\documentclass[osajnl,twocolumn,showpacs,superscriptaddress,10pt]{revtex4-1} 
\usepackage{amsmath,amssymb,graphicx,color}

\begin{document}

\title{Direct fiber excitation with a digitally controlled solid state laser source}

\author{Robert Br\"{u}ning}
\affiliation{Institute of Applied Optics, Abbe Center of Photonics, Friedrich Schiller University Jena, Fröbelstieg 1, D-07743 Jena, Germany}

\author{Sandile Ngcobo}
\affiliation{Council for Scientific and Industrial Research, National Laser Center, P.O. Box 395, Pretoria 0001, South Africa}

\author{Michael Duparr\'{e}}
\affiliation{Institute of Applied Optics, Abbe Center of Photonics, Friedrich Schiller University Jena, Fröbelstieg 1, D-07743 Jena, Germany}

\author{Andrew Forbes}\email{Corresponding author: aforbes1@csir.co.za}
\affiliation{Council for Scientific and Industrial Research, National Laser Center, P.O. Box 395, Pretoria 0001, South Africa}
\affiliation{School of Physics, University of Witwatersrand, Private Bag X3, Johannesburg 2030, South Africa}

\begin{abstract}
Mode division multiplexing has been mooted as a future technology to address the impending data crunch of existing fiber networks. Present demonstrations delineate the light source from the mode creation steps, potentially inhibiting integrated solutions.  Here we demonstrate an integrated mode generating source in the form of a digitally controlled solid state laser with an intra-cavity spatial light modulator.  In our proof-of-principle experiment we create fiber modes on demand and couple them directly into a few-mode fiber, where after transmission they are decoupled by modal decomposition.  This is the first demonstration of a single source for encoding information into the spatial modes of light.   
\end{abstract}


\maketitle 
Mode division multiplexing (MDM) is a promising solution to increase the data capacity in optical fiber systems, which are currently reaching their limits imposed by nonlinear effects \cite{Richardson2010,Richardson2013}. The building blocks of such a technology require mode multiplexing, mode transfer over fibers (or free space) and finally mode detection. Significant work has been done recently in this field to address the individual components \cite{Carpenter2012a,Bai2012,Flamm2013,Leon-Saval2014,Ho2014,Carpenter2014,Luo2014} and has resulted in demonstration of 73.7 Tb/s transmission \cite{Sleiffer2014}. However, little attention has been devoted to integrated multiplexing solutions, yet it is clear that this is a requirement for the mooted technology to be realized. In particular, to the best of our knowledge, the light source has been externally modulated prior to the fiber injection or free space transmission \cite{AlAmin2011,Koebele2011,Wang2012,Bozinovic2013,Shwartz2013}. Yet recently the concept of a digital laser was introduced for the on-demand creation of laser modes, having been demonstrated with solid state \cite{Ngcobo2013c}, and more recently fiber lasers \cite{Jung2014}. In this work we create fiber modes on demand directly from the source.  We make use of a solid state laser source with an intra-cavity spatial light modulator for the real-time mode creation.  The modes are coupled directly into a few mode fiber and transported to a mode detector.  This work represents the first proof-of-principle experiment to integrate the source with the mode creation step, and will be relevant to future mode division multiplexing demonstrations.  We use our setup to create single and superpositions of LP fiber modes and show both high fidelity in the creation ($>92$\%) and detection ($>90$\%) steps.\par

\begin{figure*}[tp]
\includegraphics[width=\textwidth]{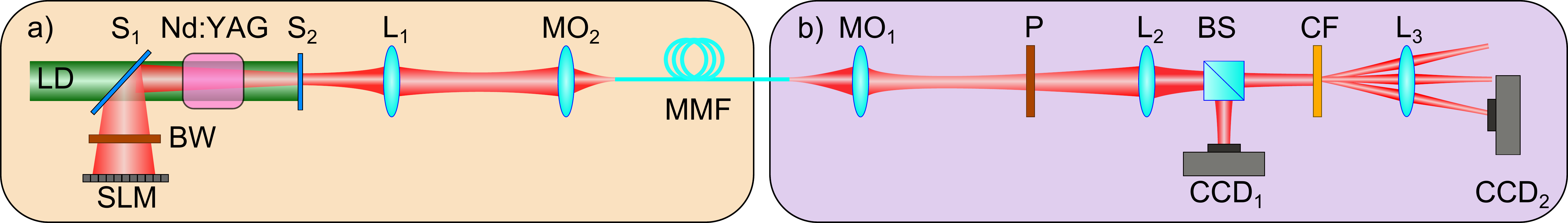}
\caption{Scheme of the experimental setup for: a) the creation, and b) the detection of fiber modes. SLM, spatial light modulator; BW, Brewster window; LD, laser diode; S$_1$ high reflective mirror, S$_2$ output coupler, Nd:YAG, gain medium; L$_1$, lens (f=200mm); MO$_{1,2}$, Microscopic objective f=10mm; L$_2$, lens (f=400mm); P, polarizer; BS, beam splitter; CF, correlation filter; L$_3$ Fourier lens (f=200mm); CCD$_{1,2}$, charge coupled device }
\label{fig.setup}
\end{figure*}

Our integrated source comprises a diode-pumped solid state (DPSS) laser where the transverse modes are controlled by an intra-cavity spatial light modulator (SLM) \cite{Ngcobo2013c}.  By programming appropriate computer generated holograms on the SLM, the desired modes can be created in the source and injected directly into the fiber delivery system. Conceptually the laser and fiber may be viewed as a single device, for example, if implemented with the recently outlined digitally controlled fiber laser \cite{Jung2014}. The concept was implemented with an electrically addressed phase-only SLM (Hamamatsu, LCOS-SLM X110468E) with pixel pitch of $20 \mu$m and reflectivity of $>90\%$ on the vertical polarization as the back reflector of our laser. The 1\% doped Nd:YAG crystal (4 mm diameter, 30 mm length) was placed in an L-shaped cavity with a 60\% output coupler (no curvature).  The output power was varied by the adjustment of the diode pump (Jenoptik JOLD 75 CPXF 2P W) power, while the desired mode size and type was selected by complex amplitude modulation on the SLM \cite{Arrizon2003b} and enables the creation of adapted mode fields to the fiber.\par
\begin{figure}[tb]
\includegraphics[width=\columnwidth]{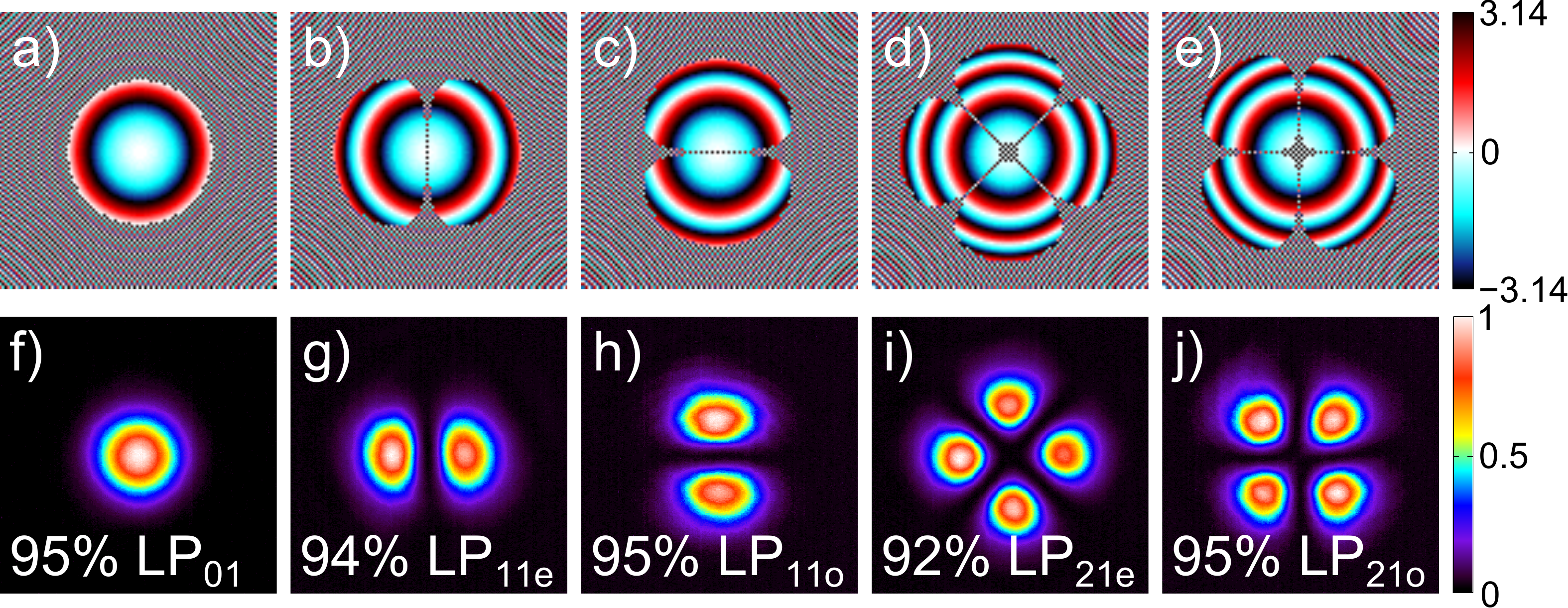}
\caption{The phase pattern of the intracavity SLM to generate the desired LP modes: a) LP$_{00}$, b) LP$_{01e}$, c) LP$_{01o}$, d) LP$_{02e}$, e) LP$_{02o}$ and the resulting intensity distributions f)-j). Insets denote the mode purity of the desired mode.}
\label{fig.input}
\end{figure}
A principle scheme of the setup for the mode generation and fiber injection is depicted in the Fig.$\,$\ref{fig.setup}a). For the first proof of principle a free space link between digital laser and optical fiber was introduced by a telescopic 4f setup with a 20$\times$ de-magnification to match the generated mode to the fiber core diameter. In principle this is not a requirement for the scheme, for example, if the digital laser source was customised to produced the desired mode sizes. The investigated fiber was a standard step index large mode area fiber (Nufern LMA-GDF-25/250-M) with a core diameter of 25 $\mu$m and a numerical aperture of 0.065. This results in a V parameter of 4.72 and the existence of six guided modes at the operation wavelength of 1064 nm.
After the transmission through the fiber the output field was 40$\times$ magnified by a second telescopic 4f setup to investigate its modal content using mode-specific match filters, otherwise known as correlation filters\cite{Kaiser2009a}. This part of the experimental setup is sketched in Fig.$\,$\ref{fig.setup}b). \par

Considering the low numerical aperture and the cylinder symmetry of the investigated fiber, their eigenstates can be assumed as linear polarized (LP) modes and each field distribution $U(\textbf{r})$ can be expressed as a weighted superposition of those
$U(\textbf{r})=\sum_{l,p}c_{lp}\varPsi_{lp}(\textbf{r})$,
where $l$ defines the azimuthal and $p$ the radial order of the modes $\varPsi_{lp}(\textbf{r})$. The modal expansion coefficients $c_{lp}=\varrho_{lp}\exp{(i\varphi_{lp})}$, with $\varrho_{lp}$ the modal weight and $\varphi_{lp}$ the modal phase, are defined by the inner product relation $c_{lp}=<U(\textbf{r}),\varPsi_{lp}(\textbf{r})>$ between the optical field $U(\textbf{r})$ and the corresponding mode $\varPsi_{lp}(\textbf{r})$. It should be noted that higher azimuthal orders are twofold degenerate. In this case it is common to distinguish between even and odd modes, whose azimuthal function is given by a cosine or sine dependency respectively, resulting in a petal like structure of the corresponding intensity distribution. The radial dependency is given by $l$th order Bessel- or modified Bessel-functions of the first kind, in the core or the cladding respectively \cite{Snyder1996}.

Good approximations of the fibers modes are given by Laguerre-Gaussian (LG) modes, which show the same principle cylindrical symmetry as the fiber modes, and are easily generated in the digital laser by applying the amplitude and phase encoded holograms, depicted in Fig.$\,$\ref{fig.input} a)-e).  
There a circular aperture and high loss lines, corresponding to the zeros of the optical field, are introduced to force the laser into the desired mode. Inside the circular aperture the SLM displays the phase function of a lens to build a stable resonator configuration. By adapting the aperture diameter and the curvature of the lens function the beam size of the out-coupled mode is controllable. This enables the generation of LG modes with the a specific beam width so as to maximise the purity of the LP mode of the fiber. The outer region of the aperture and the high loss lines are realized by a checkerboard pattern, where neighboring pixels have a phase difference of $\pi$ to get a deconstructive interference between them, which results in vanishing amplitude of the field in these regions \cite{Arrizon2003b}. This approach enabled us to produce good approximation of the LP modes with the digital laser by a judicious choice of the LG parameters.\par

The evaluation of the mode purity is based on the determination of the complex-valued modal expansion coefficients $c_{lp}$ by an all-optically realization of the inner product relation between the optical field and the desired mode distribution. For that we illuminate the correlation filter (CF), which is a computer generated hologram with the complex conjugated mode fields as the transmission functions, and perform a subsequent optical Fourier-transformation of the diffracted light by a lens in 2f configuration. This yields $I_{lp}=\varrho_{lp}^2$ as on-axis intensity in the far field and determines the modal power spectrum of the beam. The mutual phase difference $\varphi_{lp}$ of the modes can be determined in analogy using a transmission function composed of a superposition of the mode and a reference field. To achieve access to the modal information in real time the different transmission functions are multiplied with specific gratings and implemented in one correlation filter, which results in a spatial separation of the corresponding on-axis positions in the far field and allows the instant determination of all modal coefficients \cite{Kaiser2009a}. Thus our technique provides for both the real-time generation of the desired modes, and the real-time detection of them. It should be noted that if the correlation filter were fabricated onto the exit face of the fiber then the entire system would be a single integrate device.

\begin{figure}
\includegraphics[width=\columnwidth]{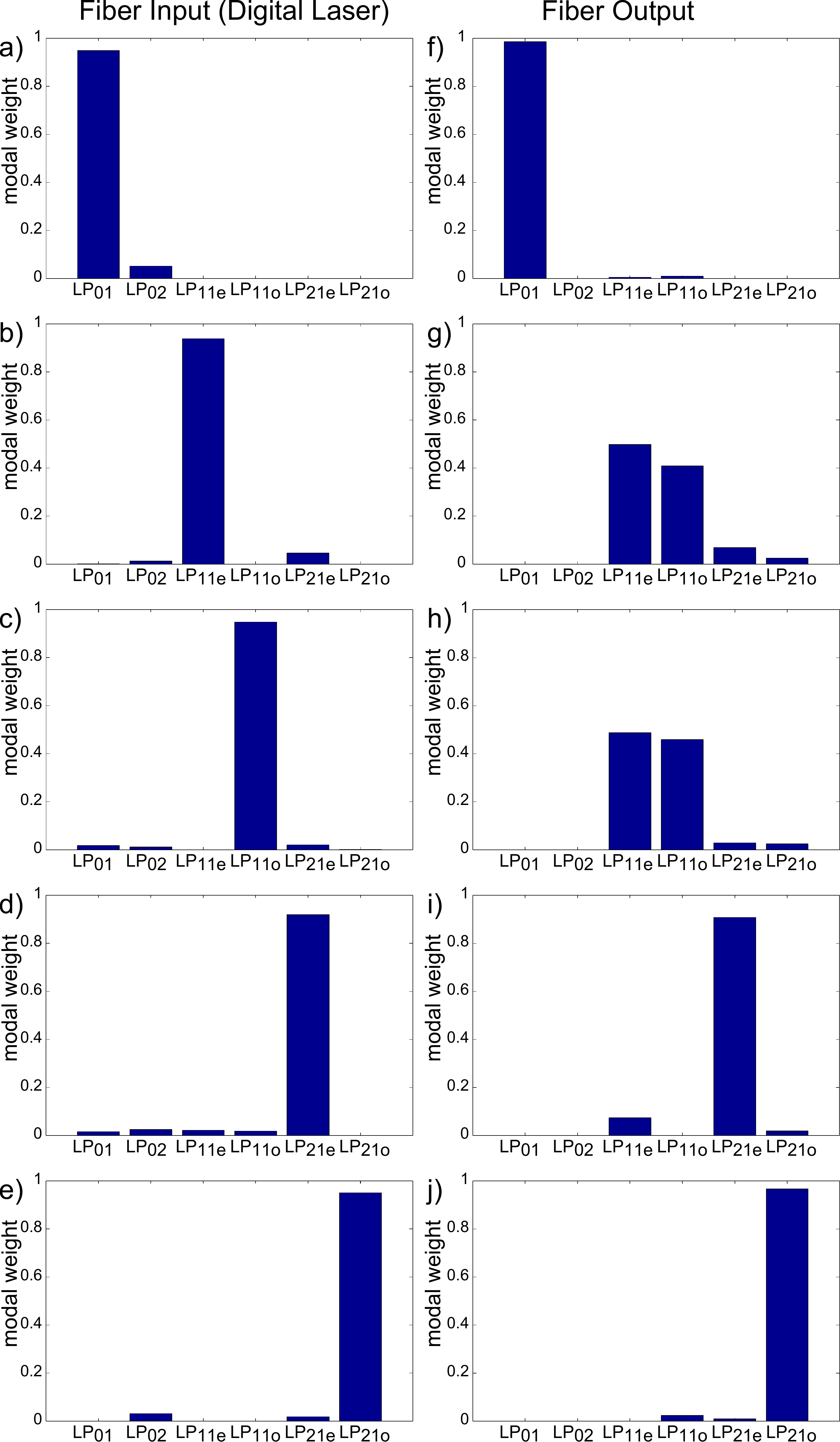}
\caption{Model power spectrum of the generated a)-e) and transmitted field f)-j) for the different created Laguerre-Gaussian modes, LG$_{00}$ for a) and f), LG$_{01e}$ for b) and g), LG$_{01o}$ for c) and h), LG$_{02e}$ for d) and i), LG$_{02o}$ for e) and j), in terms of the LP modes support by the fiber. All spectra are normalized to unity power}
\label{fig.spektrum}
\end{figure}

To evaluate the capability of our single source multiplexing scheme we first investigate the mode purity of the input signal generated by the digital laser in terms of the supported fiber modes. For that we create, with respect to the corresponding LP modes, the scale optimized mode approximation, their intensity distributions are depicted in Fig.$\,$\ref{fig.input} f)-j), and project them, by using the CFM, into the set of supported fiber modes. Conceptually the measurement scheme, the evaluation of the inner product relation, of the CFM is equivalent to the coupling into the fiber itself and provides access to the LP mode spectrum, which is excitable with the created field. Figure \ref{fig.spektrum} a)-e) shows the normalized mode spectrum in terms of the guided LP modes. It can be seen that the generated fields of the digital laser fits very well with the desired mode of the fiber. This demonstrates that the digital laser is able to address signals selective to the different modes of the fiber without generating a significant amount of cross-talk among them in the multiplexing step. Since this LP mode set is not complete in terms of describing optical free space fields, the normalization considers only the fraction of energy that can be coupled into the fiber.\par

\begin{figure}[tp]
\includegraphics[width=\columnwidth]{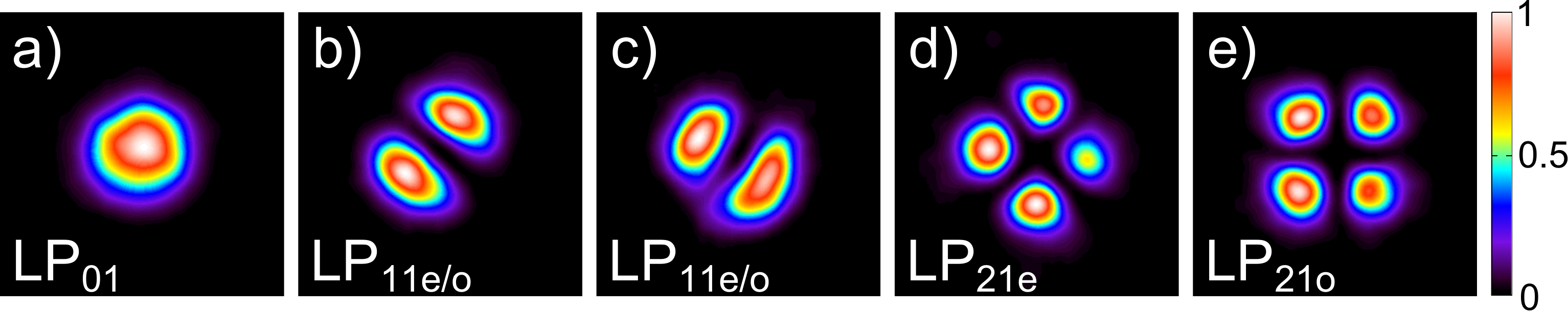}
\caption{The intensity distributions at the output of the optical fiber, the main contributed modes are denoted.}
\label{fig.output}
\end{figure}

In a second step we inject the mode fields of the digital laser into a short piece of the optical fiber ($\approx$30cm). The short length was chosen to avoid possible distortion effects on the signal, which could be caused by the fiber and not by the digital laser. Then the output signal of the fiber was investigated by modal decomposition to determine the mode purity of the transmitted signal. The intensity distributions at the fiber output are depicted in Fig.$\,$\ref{fig.output} and are commensurate to the structure of the input fields, see Fig.$\,$\ref{fig.input}. For the inserted LP$_{11e}$ (Fig.$\,$\ref{fig.output} b)) and LP$_{11o}$ (Fig.$\,$\ref{fig.output} c)) modes it is remarkable that the two petal structure is in principle preserved but rotated by an angle of approximately 45$^\circ$. In terms of the given modal set, which has an orientation of the petal structure in vertical and horizontal direction, the output filed is given by superposition of the LP$_{11e}$ and LP$_{11o}$ modes and indicates some mode coupling effects during the propagation through the fiber. The rotation angle between input and output field was changeable by slightly stressing the fiber, which also indicates that the mode coupling is a propagation effect through the fiber and not induced by the coupling of the digital laser into the fiber. \par
\begin{figure*}
\includegraphics[width=\textwidth]{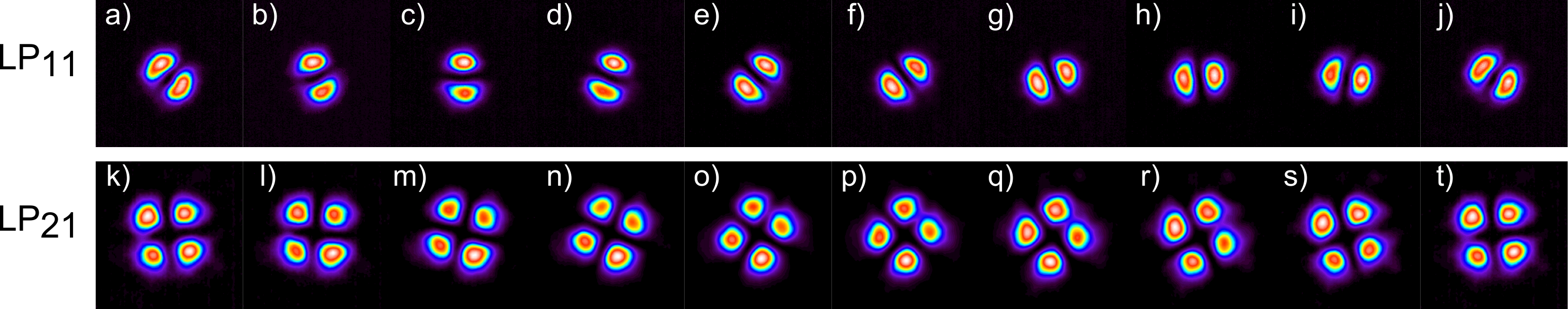}
\caption{Dynamical orientation change of the petal like modes LP$_{11}$ a)-j) and LP$_{21}$ k)-t), by 180 respectively 90 degree. The rotation of the transmitted intensity in respect to the phase pattern of the intracavity SLM and the related normalized modal power spectrum can be seen in Media 1 for the LP$_{11}$ mode and Media 2 for the LP$_{21}$ mode}
\label{fig.rotation}
\end{figure*}

The result of the modal decomposition on the fiber output are shown in Fig.$\,$\ref{fig.spektrum} f)-j), and confirm quantitatively the high modal purity of the transmitted signals. In all cases, considering the sum of the degenerated modes for the LP$_{11}$ like output, a high mode purity about 90\% of the transmitted beam was achievable. In comparison with the spectra of the input signal (Fig.$\,$\ref{fig.spektrum} a)-e)) the mode purity change, for each specific mode, is below 4\% and confirm the ability to address signals into specific fiber modes with the digital laser.

A second advantage of the digital laser, beside the generation of adapted mode fields, is the possibility to change them dynamically only limited by the image refresh time of the SLM. Figure \ref{fig.rotation} shows two examples of the creation and injection of different weighted superpositions of the petal like modes LP$_{11}$ Fig.$\,$ \ref{fig.rotation} a)-j) and LP$_{21}$ Fig.$\,$ \ref{fig.rotation} k)-t), which results in a rotation of their orientation. The dynamical change of the intensity distribution, together with the corresponding phase pattern of the intra-cavity SLM and the resulting mode spectra can be seen in media 1 for the LP$_{11}$ mode and in media 2 for the LP$_{21}$ mode. This capability to change the modal spectrum at the input can also be used to compensate distortion effects of the fiber, for example, the observed coupling between the LP$_{11e}$ and LP$_{11o}$ mode, which results in the rotated output pattern. By exciting a suitable superposition of these modes at the input of the fiber, the coupling can be precompensated to achieve the desired pure LP$_{11e}$ and LP$_{11o}$ modes at the output, see Fig.$\,$\ref{fig.rotation} c) and h).\par

In conclusion, we have realized the addressing of specific modes in an optical few mode fiber directly from the laser source. This avoids the usage of inefficient extra-cavity beam shaping techniques or multiple source setups. Additionally, the capability to compensate fiber introduced cross talk was shown by a suitable adaption of the input mode spectrum. Further attention should be spent on directly coupled solutions, which should increase the mode purity by avoiding aberration effects introduced by the telescopic setups used.

\end{document}